# Implications of the universal scaling relation in high-temperature superconductors

A. Rosencwaig*

ARIST, Danville, CA 94506

## **ABSTRACT**

The recently discovered universal scaling relation between the superconducting density and the transition temperature in high-temperature superconductors appears to indicate that those normal state carriers that are undergoing a superconducting transition at $T \cong T_c$, are experiencing critical dissipation and acting as preformed pairs.

Recently, Homes *et al* have reported that a simple scaling relation between the superconducting transition temperature and the superfluid density universally holds for all high-temperature superconductors, regardless of doping level, crystal structure and type of disorder, nature of dopant and direction.[1] Using a dimensional analysis, Zaanen has proposed that this scaling relation implies that the normal states in the high-temperature superconductors are extremely, or even critically, dissipative.[2] The Homes relation is given by,

$$\rho_s = \eta \sigma_{dc} T_c \qquad (1)$$

where $\rho_s$ is the superfluid density and $\sigma_{dc}$ is the dc conductivity at a temperature $T \cong T_c$, where $T_c$ is the transition temperature. Homes *et al* have determined that $\eta = 120 \pm 25$ $\Omega cm^{-1} K^{-1}$ when $\rho_s$ is in $cm^{-2}$ and $\sigma_{dc}$ is in $\Omega^{-1} cm^{-1}$.

Now,

$$\rho_s = \frac{c^2}{\lambda^2} = \frac{4\pi e^2 n_s}{m_s} \qquad (2)$$

where $c$ is the velocity of light, $\lambda$ is the penetration depth and $m_s$, $n_s$ and $e$ are the mass (or effective mass), the density and the charge, respectively, of the carriers in the superconducting state at $T = 0$. The normal state dc conductivity in the $CuO_2$ planes, where coherent transport governed by scattering physics applies, is given by,

$$\sigma_{dc} = \frac{ne^2 \tau}{m} = \frac{ne^2 \Lambda}{mv} \qquad (3)$$

where $n$ is the carrier density, $\tau$ is the relaxation time, $m$ is the carrier effective mass, $\Lambda$ is the carrier mean free path or scattering length and $v$ is the average carrier velocity. At $T \cong T_c$, the normal conductivity has already significantly increased and thus is dominated by those carriers (density $n' \leq n$) that are undergoing the superconducting transition. Thus $n' = n_s$. Since $n'$ can be $< n$, this is consistent with Tanner's rule[3] which states that in high-$T_c$ cuprates $n_s$ tends to be $\approx \frac{1}{4} n$. Thus,

$$\sigma_{dc}(T \approx T_c) = \frac{n' e^2 \Lambda}{m' v} = \frac{n_s e^2 \Lambda}{m' v} \qquad (4)$$

where $m'$ is the effective mass of those normal state carriers that are undergoing a superconducting transition at $T \cong T_c$.

But,

$$m'v = \frac{2\pi\hbar}{\lambda_T} \quad (5)$$

where $\hbar$ is the Planck constant and $\lambda_T$ is the de Broglie thermal wavelength for the quasiparticles that are undergoing the superconducting transistion at $T \cong T_c$. Therefore, substituting both Eqn. (5), and $n_s$ from Eqn. (4), into Eqn. (1), we get,

$$\rho_s = \frac{8\pi^2\hbar}{m'\Lambda\lambda_T}\sigma_{dc} = \frac{8\pi^2\hbar}{m_s\lambda_T^2}\left(\frac{\lambda_T}{\Lambda}\right)\sigma_{dc} \quad (6)$$

But,

$$\lambda_T^2 = \frac{2\pi\hbar^2}{m'kT_c} \quad (7)$$

where $k$ is the Boltzmann constant. Thus,

$$\rho_s = \frac{4\pi k}{\hbar}\left(\frac{\lambda_T}{\Lambda}\right)\left(\frac{m'}{m_s}\right)\sigma_{dc}T_c \quad (8)$$

The proportionality factor $\eta$ in Eqn. (1) is then,

$$\eta = \frac{4\pi k}{\hbar}\left(\frac{\lambda_T}{\Lambda}\right)\left(\frac{m'}{m_s}\right) \quad (9)$$

This is the expression for $\eta$ when $\rho_s$ is in units of s$^{-2}$ and $\sigma_{dc}$ is in units of s$^{-1}$ giving units of $\eta$ in s$^{-1}$K$^{-1}$. To compare theory with experiment, we must set $\rho_s$ and $\sigma_{dc}$ to be in units of cm$^{-2}$ and $\Omega^{-1}$cm$^{-1}$ respectively. To do this, we divide the right-hand side of Eqn. (9) by $c^2$ and the left-hand side by $\Omega c$, which results in,

$$\eta = \frac{4\pi k}{\hbar c}\left(\frac{\lambda_T}{\Lambda}\right)\left(\frac{m'}{m_s}\right) \quad (10)$$

$$= 55\left(\frac{\lambda_T}{\Lambda}\right)\left(\frac{m'}{m_s}\right) \; \Omega\text{cm}^{-1}\text{K}^{-1}$$

Since the experimental value for $\eta$ is $120 \pm 25$ $\Omega$ cm$^{-1}$K$^{-1}$,[1]

$$\left(\frac{\lambda_T}{\Lambda}\right)\left(\frac{m'}{m_s}\right) \approx 2 \qquad (11)$$

One way for Eqn. (12) to be true is if both $\Lambda \approx \lambda_T$ and $m' \approx 2m_s$. The first condition is similar to Zaanen's suggestion that the normal carriers at $T \cong T_c$ undergo critical dissipation,[2] although here this condition may be limited to only those normal carriers that are undergoing the superconducting transition at $T \cong T_c$. The second condition implies that the normal carriers undergoing the superconducting transition at $T \cong T_c$ are acting as preformed pairs. Again, this does not necessarily mean that all the normal carriers are preformed pairs, only that those carriers undergoing the superconducting transition at $T \cong T_c$ act as preformed pairs.

The fact that the Homes relation also holds for the c-axis as well is surprising given that the normal state transport in this direction is generally governed by hopping rather than scattering physics. A possible explanation may be that at $T \cong T_c$ preformed pairs begin to form along the c-axis and that these preformed pairs undergo coherent transport along the c-axis that is governed by scattering physics, albeit with critical dissipation with appropriately larger effective mass $m'$ and smaller thermal wavelength $\lambda_T$ and scattering length $\Lambda$.

## ACKNOWLEDGMENTS

I would like to acknowledge helpful discussions with M. Greven and C.C. Homes.